\documentclass[final]{siamltex}
\usepackage{amsfonts}
\usepackage{amssymb}
\usepackage{latexsym}
\usepackage{ifthen}

\newcommand{\ra}{\rightarrow}

\newcommand{\tx}{\textrm}

\newtheorem{thm}{Theorem}
\newtheorem{lem}{Lemma}
\newtheorem{prop}{Proposition}
\newtheorem{cor}{Corollary}

\title{The predictable degree property, column reducedness, and minimality in multidimensional convolutional coding}

\author{Vakhtang Lomadze\thanks{A. Razmadze Mathematical Institute, Mathematics Department of I. Javakhishvili State University, Georgia ({\tt vakhtang.lomadze@tsu.ge}).}
        }
\begin{document}

\maketitle
\begin{center}\footnotesize{\textsl{ {
}}}
\end{center}

\begin{abstract}
Higher-dimensional analogs of  the predictable degree property and column reducedness are defined, and it is proved that
the two properties are equivalent. It is  shown that every multidimensional convolutional code has, what is called,  a minimal
reduced polynomial resolution. It is uniquely determined (up to isomorphism) and  leads to a number of important
integer invariants of the code generalizing classical Forney's indices.
\end{abstract}

\begin{keywords}
Convolutional code, filtered module, PD property, polynomial resolution, reduced polynomial resolution, graded module, column degree table, Forney table.
\end{keywords}

\pagestyle{myheadings}
\thispagestyle{plain}

\section{Introduction}

Multidimensional convolutional codes are natural generalizations of classical (one-dimensional)
convolutional codes and are used to transmit
multidimensional data. They have been studied quite a
bit in the literature and we refer the reader
to     Fornasini and Valcher \cite{FV},  Valcher and Fornasini \cite{VF}, Weiner \cite{W}
 and more recent works Charoenlarpnopparut \cite{C}, Gluesing-Luerssen et al. \cite{GRW}, Jangisarakul and Charoenlarpnopparut \cite{JC}, Kitchens \cite{K}, Napp Avelli et al.
\cite{NPP1,NPP2}, Zerz \cite{Z}.

In this article, we would like to offer a new view-point on some
 fundamental issues of algebraic character related to multidimensional convolutional codes.

Throughout,  $\mathbb{F}$ is an arbitrary (finite) field, $n$  a fixed positive integer,
and $D=(D_1,\ldots ,D_n)$  a sequence of indeterminates. We let $S=\mathbb{F}[D]$. For
every  $d\in \mathbb{Z}$, we shall write $S_{\leq d}$ to denote the space of polynomials of degree
$\leq d$.

Following Weiner \cite{W} and other authors, by a  convolutional code of length $q$ we  mean
  a submodule  of $S^q$.
If $C$ is a convolutional code of length  $q$, then, for each $d\geq 0$, define
$
C_{\leq d}=C\cap S^q_{\leq d}
$
to be the space of codewords of degree $\leq d$.

In dimension 1, a submodule $C\subseteq S^q$ is free, and it is possible therefore
 to represent it via a full column rank polynomial matrix. In other words, there exist an integer $p$ and a polynomial matrix
$G\in S^{q\times p}$ such that
$$
0\ra S^p\stackrel{G}{\ra} C\ra 0
$$
is an exact sequence.
A higher-dimensional counterpart of this well-known fact is
Hilbert's syzygy theorem, according to  which there exists  $1\leq l \leq n$, and there exist integers
$p_1, \ldots , p_l \geq 1$ and polynomial matrices $G_1, \ldots , G_l$ of sizes
$q\times p_1, \ldots , p_l\times p_{l-1}$, respectively, such that the sequence
$$
0\ra S^{p_l}\stackrel{G_l}{\ra} S^{p_{l-1}}\ra \cdots \ra S^{p_1}\stackrel{G_1}{\ra} C\ra 0
$$
is exact.  This celebrated theorem suggests  to define higher-dimensional analogs of classical full column rank polynomial matrices
as sequences $(G_l, \ldots , G_1)$ of polynomial matrices satisfying the above exactness condition.

Starting with this idea, we  shall define higher-dimensional analogs of  the predictable degree property and column reducedness,
and prove that
the two properties are equivalent. We shall see that every multidimensional convolutional code has, what we call, a minimal
reduced polynomial resolution. It is uniquely determined (up to isomorphism) and  provides a number of important
integer invariants of the code that generalize classical Forney's indices.

 Recall that the degree of a
 column $f$ with entries in $S$ is defined to be the maximum of the degrees of the components of $f$; it is denoted by $\deg(f)$.
We shall need {\em relative} degrees as well.
Any function $[1, p] \ra \mathbb{Z}_+$ will be referred to as a twisting (or degree) function of length $p$. If $a$ is a twisting function
of length $p$, then, for
 $f\in S^p$, we set
$$\deg_a(f)=\max_i\{a(i)+\deg(f_i)\}.
$$
(If $f$ is zero, then $\deg_a(f)=-\infty$.) Notice that $\deg(f)=\deg_0(f)$.

 Without loss of generality, we shall always consider polynomial matrices
with no zero column. For such matrices, we can define (column) degree functions.
The  degree function of  a polynomial matrix $G\in S^{p\times q}$, denoted by $\deg(G)$, is the
function that assigns to every $j\in [1, q]$ the degree of the $j$-th column of $G$. More generally, if
 $a$ is a twisting function
of length $p$, we define the $a$-degree function $\deg_a(G):[1,q] \ra \mathbb{Z}_+$ to be the function
that assigns to every $j\in [1, r]$ the $a$-degree of the $j$-th column of $G$.
This is a degree function of length $q$.

By a  polynomial complex of length $l$, we shall mean a sequence $(G_l,\ldots , G_1)$ of polynomial matrices,
such that the products
$G_1G_2,\ \ldots ,\ G_{l-1}G_l$ (are defined and) are zero.
The size is defined to be  $q\times (p_l, \ldots ,p_1)$, where $q$ is the row number of $G_1$ and $p_1, \ldots , p_l$ are
 the column numbers of $G_1, \ldots , G_l$.
We say that $(G_l,\ldots , G_1)$ is a polynomial resolution  if
$$
0\ra S^{p_l}\stackrel{G_l}{\ra} S^{p_{l-1}}\ra \cdots \ra S^{p_1}\stackrel{G_1}{\ra} S^q.
$$
is an exact sequence.

If $C$ is a convolutional code, then, as already said, Hilbert's syzygy theorem guarantees existence of
 a polynomial resolution $(G_l,\ldots , G_1)$ such that
$\tx{Im}(G_1)=C$. The number $l$, called the homological dimension of $C$, is an important integer invariant of $C$. This number measures the complexity of $C$ and indicates how far is $C$ from being free.
 Free convolutional codes
are exactly convolutional codes of homological dimension 1.

  The point of convolutional codes is that they admit a  natural
 {\em homogenization}, and this permits us to study them using the method of graded modules. We introduce  an extra ("homogenizing") indeterminate $D_0$, and  define $T=\mathbb{F}[D_0,D]$. Given an integer function
 $a:[1,p]\ra \mathbb{Z}$, we shall write $D_0^a$ for the diagonal matrix with $D_0^{a(1)}, \ldots , D_0^{a(p)}$ on the diagonal.

This article has much overlaps with \cite{L2}. The relevant results from \cite{L2} are reproduced in a rather sketched form. The main new contribution is Theorem 1, which generalizes Forney's classical theorem stating that
a full column rank polynomial matrix has the predictable degree property if and only if its leading coefficient matrix
has full column rank.

\section{Filtered modules, and the PD property}

Let $M$ be a module over $S$.
A filtration on $M$ is an ascending chain
 $$M_{\leq 0} \subseteq M_{\leq 1} \subseteq  M_{\leq 2} \subseteq \cdots $$
 of linear subspaces of $M$
such that $$M=\bigcup M_{\leq d}\ \ \ \tx{and}\ \ \
  D_kM_{\leq d} \subseteq M_{\leq d+1}\ \forall k,d.
$$
A module equipped with a filtration is called a filtered module.

A twisting function $a:[1, p]\ra \mathbb{Z}_+$ determines on
$S^p$  a  filtration  consisting of the spaces
$$
S^p[-a]_{\leq d}=\{f\in S^p |\ \deg_a(f)\leq d \} \ \ \ (d\geq 0).
$$
The module $S^p$ equipped with this filtration is denoted by $S^p[-a]$.
Given a submodule  $C\subset S^p$, we shall write $C[-a]$ to denote
the module $C$ together with the filtration induced from $S^p[-a]$, that is,
$$
C[-a]_{\leq d}=C\cap S^p[-a]_{\leq d}.
$$

A homomorphism of filtered modules $(M, (M_{\leq d}))\ra (N, (N_{\leq d}))$ is a homomorphism $u:M\ra N$ such that
$$\forall d\geq 0,\ \ \ u(M_{\leq d})\subseteq N_{\leq d}.$$

{\em Example}. If $a: [1, p]\ra \mathbb{Z}_+$ and $b: [1, q]\ra \mathbb{Z}_+$ are two  functions, then
$$
\tx{Hom}(S^q[-b],S^p[-a])= \ \ \ \ \ \ \ \ \ \ \ \ \ \ \ \ \ \ \ \ \ \ \ \ \ \ \ \ \ \ \ \ \ \ \ \ \ \ \ \
$$
$$ \ \ \ \ \ \ \ \ \ \ \ \ \ \ \ \ \ \ \ \ \ \ \ \{(g_{ij})\in S^{p\times q}|\ \deg(g_{ij})\leq b(j)-a(i)\}.
$$

Certainly, filtered modules and their homomorphisms form a category.
Consequently, we may speak,  in particular, about isomorphisms between filtered modules.

\begin{lem} Let $a: [1, p]\ra \mathbb{Z}_+$ and $b: [1, q]\ra \mathbb{Z}_+$ be two twisting functions. If
$$
  S^p[-a] \ \simeq \ S^q[-b],
$$
then \ $p=q$ and $a=b$ (up to permutation).
\end{lem}

{\em Proof}.  That $p=q$ is obvious. Proving the second equality, we may assume that $a$ and $b$ are increasing functions.
Suppose that $a\neq b$, and let $i$ be the smallest number such that $a(i)\neq b(i)$. Say that $a(i)>b(i)$. Letting $d=a(i)$, we have:
$$
S^p[-a]_{\leq d} \ \simeq \ S^p[-b]_{\leq d}.
$$
But the left side here is equal to
$$
S_{\leq a(1)-d}\oplus \cdots \oplus S_{\leq a(i-1)-d}\oplus \mathbb{F} \oplus \cdots
$$
and the right side is
$$
S_{\leq b(1)-d}\oplus \cdots \oplus S_{\leq b(i-1)-d}.
$$
We get a contradiction.
 $\quad\Box$

The category of filtered modules is not  abelian.  Nevertheless, we may speak about  exact
sequences in it. Call
a complex of filtered modules
$$
(F_l,(F_{l,\leq d}))\stackrel{\delta_l}{\ra} \cdots \stackrel{\delta_2}{\ra}
  (F_1,(F_{1,\leq d}))
\stackrel{\delta_1}{\ra} (F_0,(F_{0,\leq d}))
$$
 exact if the sequence
$$
F_{l,\leq d}\stackrel{\delta_l}{\ra} \cdots \stackrel{\delta_2}{\ra}
  F_{1,\leq d}
\stackrel{\delta_1}{\ra} F_{0,\leq d}
$$
is exact for all $d\geq 0$.

\begin{lem} If a complex of filtered modules
$$
(F_l,(F_{l,\leq d}))\stackrel{\delta_l}{\ra} \cdots \stackrel{\delta_2}{\ra}
  (F_1,(F_{1,\leq d}))
\stackrel{\delta_1}{\ra} (F_0,(F_{0,\leq d}))
$$
is  exact, then the complex  of modules
$$
F_l\stackrel{\delta_l}{\ra} \cdots \stackrel{\delta_2}{\ra}
  F_1
\stackrel{\delta_1}{\ra} F_0
$$
also is exact.
\end{lem}

{\em Proof}. This is obvious  because $F_i=\lim_{d\ra 0}F_{i,\leq d}$ (and because the direct limit functor is an exact functor).
 $\quad\Box$

{\bf Definition.}\  Let $G=(G_l,\ldots , G_1)$ be a polynomial complex, and
let $q\times (p_l, \ldots ,p_1)$ be its size.
 Define the degree  functions
 $$
 a_i:[1,p_i]\ra \mathbb{Z}_+,\ \ i=1,\ldots , l
 $$
 recursively as follows. Put
$a_0=0$, and if $a_i$ is defined,  define $a_{i+1}$ to  be
$$
a_{i+1}=\deg_{a_i}(G_{i+1}).
$$
Call $(a_l, \ldots ,a_1)$ the column degree table of $G$.
The  polynomial complex $G$ gives rise to  the following complex of filtered modules
\begin{equation}
0\ra S^{p_l}[-a_l]\stackrel{G_l}{\ra}  \cdots \stackrel{G_2}{\ra} S^{p_1}[-a_1]
\stackrel{G_1}{\ra} C[0]\ra 0.
\end{equation}
We say that $G$
  has the PD (predictable degree) property if this complex is exact.

The following example justifies the above definition.

{\em Example}. Assume $n=1$. Following Forney \cite{F},
a polynomial matrix $G\in S^{q\times p}$ is said to have  the PD property if, for every $f\in S^p$,
$$
\deg(Gf)=\deg_a(f),
$$
where $a$ is the column degree function  of $G$.
 For each $d\geq 0$, $G$ determines a linear map
$$
S^p[-a]_{\leq d}\ra C_{\leq d},
$$
and it is easily seen that $G$ has the PD property in the sense of Forney if and only if all these linear maps are bijective.

Let $C$ be a convolutional code of length  $q$.
Define the Hilbert function of $C$ as
$$
\tx{HF}(C,d)=\dim_\mathbb{F}(C_{\leq d}), \ \ \ d\in \mathbb{Z}_+.
$$
This can be easily computed from the column degree table of a  polynomial resolution of
$C$ having the PD property. Indeed, if
 $(G_l,\ldots , G_1)$ is such a resolution and if
  $(a_l, \ldots ,a_1)$ is its column degree table, then, for each $d\geq 0$,
  \begin{equation}
0\ra S^{p_l}[-a_l]_{\leq d}\stackrel{G_l}{\ra} \cdots \stackrel{G_2}{\ra}
S^{p_1}[-a_1]_{\leq d}\stackrel{G_1}{\ra} C_{\leq d}\ra 0
\end{equation}
is an exact sequence of finite-dimensional linear spaces.
As is known, the dimension of the space of polynomials
of degree $\leq d$, where $d\in \mathbb{Z}$,  is equal to $\left(\begin{array}{c}d+n\\n\end{array}\right)$ (with the  convention that
$\left(\begin{array}{c}d+n\\n\end{array}\right)=0$ when $d\leq -1$).
It immediately follows from (2) therefore that
$$
\tx{HF}(C,d)=\sum_{i=1}^l(-1)^{i-1}\sum_j \left(\begin{array}{c}d-a_i(j)+n\\ n\end{array}\right).
$$

{\em Remarks}.  Hilbert functions  were introduced by Hilbert  in the context of graded finitely generated modules.
For one-dimensional convolutional codes, they have been defined in McEliece and Stanley \cite{MS}. Notice that
the above expression
for Hilbert function is a generalization of that given in [Corollary 3.2, 13].

\section{The homogenization and the leading term complex}

According to Lemma 2, the polynomial complexes having the PD property are polynomial resolutions
necessarily. The converse is not true, and the  goal of this section is to obtain  a test
to establish whether a polynomial resolution has the predictable degree property or not.

We need to recall the notion of graded modules.

Let $M$ be a module over $S$ or $T$. A gradation on $M$ is a sequence $$M_0,M_1,M_2,\ldots $$ of $\mathbb{F}$-linear subspaces of $M$
such that
$$
M=\bigoplus M_d \ \ \ \tx{and}\ \ \ D_kM_d\subseteq M_{d+1}\ \forall k,d.
$$
A module equipped with a gradation is called a graded module.

{\em Example}. Let $R$ be either $S$ or $T$. A twisting function $a$
of length $p$ determines on $R^p$ the gradation  consisting of the spaces
$$
R^p(-a)_d=\{f\in S^p |\ \deg(f_i)= d-a(i) \} \ \ \ (d\geq 0).
$$
The module $R^p$ equipped with this gradation will be denoted by
$R^p(-a)$.

A homomorphism of graded modules $(M, (M_{d}))\ra (N, (N_{d}))$ is a homomorphism $u:M\ra N$ such that
$$\forall d\geq 0,\ \ \ u(M_{d})\subseteq N_{d}.$$
Graded modules  form an abelian category, and one therefore has the notion of exact
sequences. It is worth noting  that
a complex of graded modules
$$
(F_l,(F_{l,d}))\stackrel{\delta_l}{\ra} \cdots \stackrel{\delta_2}{\ra}
  (F_1,(F_{1,d}))
\stackrel{\delta_1}{\ra} (F_0,(F_{0,d}))
$$
is exact if only if the sequence of linear spaces
$$
 F_{l,d}\stackrel{\delta_l}{\ra} \cdots \stackrel{\delta_2}{\ra}
  F_{1,d}
\stackrel{\delta_1}{\ra} F_{0,d}
$$
is exact for all $d\geq 0$.

The homogenization in degree $d$ is the bijective linear map
$$
S_{\leq d}\ra T_d
$$
defined by
$$f(D)\mapsto D_0^df(D/D_0).$$ (Here and below $D/D_0$ means $(D_1/D_0,\ldots , D_n/D_0)$.)

({\em Warning}: It is essential  to indicate the "$d$". For instance, the homogenization in degree 4 of the polynomial $2D_1^3D_n+1$ is
 $2D_1^3D_n+D_0^4$ and the homogenization in degree 5 is $2D_0D_1^3D_n+D_0^5$.)

Let $C\subseteq S^p$ be a convolutional code.
The homogenization $C^H$ of $C$ is defined to be
$$C^H=\bigoplus_{d\geq 0} C^H_d,$$
where $C^H_d$ is the image of $C_{\leq d}=C\cap \mathbb{F}[s]^p_{\leq d}$ under the homogenization operator
$S^p_{\leq d}\ra T^p_d$.
This is a "homogeneous convolutional code" in $T^p$.

{\bf Definition} \ Let $G=(G_l,\ldots ,G_1)$ be a polynomial complex, and let $q\times (p_l,\ldots ,p_1)$
be its size and $(a_l,\ldots ,a_1)$ the column degree.

a) The homogenization of $G$ is the sequence $G^H=(G_l^H,\ldots ,G_1^H)$ of homogeneous polynomial matrices,
where $G_k^H=D_0^{-a(k)}G_k(D/D_0)D_0^{a(k+1)}$.

b) The  leading term complex of $G$  is the sequence $G^L=(G_1^L,\ldots ,G_l^L)$ of
homogeneous polynomial matrices, where
$G_k^L$ is defined as follows. The  $(i,j)$ entry of $G_k^L$ is
the homogeneous  ($a_{k+1}(j)-a_k(i)$)-th part of the $(i, j)$ entry of $G_k$.

These two sequences  give rise respectively to the following complexes of graded modules:
\begin{equation}
0\ra T^{p_l}(-a_l)\stackrel{G_l^H}{\ra}  \cdots \stackrel{G_2^H}{\ra} T^{p_1}(-a_1)
\stackrel{G_1^H}{\ra} T^q(0)
\end{equation}
and
\begin{equation}
0\ra S^{p_l}(-a_l)\stackrel{G_l^L}{\ra}  \cdots \stackrel{G_2^L}{\ra}
S^{p_1}(-a_1)\stackrel{G_1^L}{\ra} S^q(0).
\end{equation}
We remark that
$$
G^L(D)=G^H(0,D),
$$
that is, $G^L$ is obtained from $G^H$ by replacing $D_0$ by 0.

\begin{thm} Let $G=(G_l,\ldots, G_1)$ be a polynomial complex, and let $q\times (p_l,\ldots ,p_1)$ be its size
and $(a_l,\ldots ,a_1)$ the column degree. The following three conditions are
equivalent:

(a) $G$ has the PD property;

(b) $ImG_1^H=C^H$ and the complex (3) is exact;

(c) $G$ is a polynomial resolution and the complex (4) is exact.
\end{thm}

{\em Proof}. (a)\ $ \Leftrightarrow $ \ (b) This is obvious since the complex (2)
is isomorphic to the complex
$$
0\ra T^{p_l}(-a_l)_d\stackrel{G_l^H}{\ra}  \cdots \stackrel{G_2^H}{\ra}
  T^{p_1}(-a_1)_d\stackrel{G_1^H}{\ra} C^H_d\ra 0.
$$
(b)\ $\Rightarrow $ \ (c)  We have an exact sequence
$$
0\ra T^{p_l}(-a_l)\stackrel{G_l^H}{\ra}  \cdots \stackrel{G_2^H}{\ra} T^{p_2}(-a_2)\stackrel{G_1^H}{\ra} T^q\ra T^q/C^H.
$$
We claim that $D_0$ is not a zero divisor on $T^q/C^H$.
Suppose that $u\in T^q_d$ is such that $D_0u\in C^H_{1+d}$. Then $u(1,D)\in C$.
 Since
$u(1,D)$ has degree $\leq d$, we have $u(1,D)\in C_{\leq d}$. Clearly, $u$ is the $d$-homogenization of
$u(1,D)$, and therefore belongs to $C^H_d$.
Using Corollary 1 in  \cite{L1}, we can see that the sequence
$$
0\ra T^{p_l}(-a_l)\stackrel{G_l^H}{\ra} T^{p_{l-1}}(-a_{l-1})\ra \cdots \stackrel{G_1^H}{\ra} T^q.
$$
remains exact after tensoring it by $T/D_0T=S$. Notice that replacing $D_0$ by 0 in each entry of $G_i^H$, we get $G_i^L$. This means that
$$G_i^H\otimes T/D_0T=G_i^L.$$
(b)\ $\Rightarrow $ \ (a)  Let $d\geq 0$. We have to show that the sequence
 $$
0\ra S^{p_l}[-a_l]_{\leq d}\stackrel{G_l}{\ra} \cdots \stackrel{G_2}{\ra}
  S^{p_1}[-a_1]_{\leq d}
\stackrel{G_1}{\ra} C_{\leq d}\ra 0
$$
is exact. For $1\leq i\leq l-1$, we set
$$ X_i=
Ker(S^{p_i}[-a_i]_{\leq d}\stackrel{G_i}{\ra}S^{p_{i-1}}[-a_{i-1}]_{\leq d});
$$
if $i=0$, define $X_i$ to be $C_{\leq d}$.

Take any $x\in X_i$. By the hypothesis,
there exists $y\in  S^{p_{i+1}}$ such that $G_{i+1}y=x$. If  $\deg_{a_{i+1}}(y)\leq d$, we are done.
 If $k=\deg_{a_{i+1}}(y)> d$, write
 $$y=y^L+y^{\prime}$$
 with $y^L\in S^{p_{i+1}}(-a_{i+1})_k$ and $y^\prime \in S^{p_{i+1}}[-a_{i+1}]_{\leq k-1}$.
 It is clear that $G_{i+1}^Ly^L=0$. Hence, by the hypothesis,
 $G_{i+2}^Lz=y^L$ for some   $z\in S^{p_{i+2}}(-a_{i+2})_k$.
Consider the element $y_1=y-G_{i+2}z$. It has $a_{i+1}$-degree $<k$ and $$G_{i+1}y_1=G_{i+1}y-G_{i+1}G_{i+2}z=G_{i+1}y=x.$$
Using induction, we find  that there  exists an element in $S^{p_{i+1}}[-a_{i+1}]_{\leq d}$ that goes to $x$.

The proof is complete.
 $\quad\Box$

{\bf Definition.}\
Let $G=(G_1,\ldots, G_l)$ be a polynomial complex, and let $q\times (p_l,\ldots ,p_1)$ be its size
and $(a_l,\ldots ,a_1)$ the column degree. Say that $G$ is (column) reduced if the complex (4) is exact.

We thus have the following statement.

\begin{cor}
A polynomial resolution has the PD
property if and only if it is reduced.
\end{cor}

{\em Remark}.  In dimension 1, if $G$ is a full column
rank polynomial matrix with degree function $a$, then $G^L$ is equal to the leading coefficient matrix multiplied by the diagonal matrix
with $D^{a(i)}$ on the diagonal.
For example, if
$$
G=\left[\begin{array}{cc}2D^3+D+1&D^2-10\\
D^2-5&D+4\\
3D^4+7D&D^2+1
\end{array}\right],
$$
then the column degree is equal to $(4,2)$, and we have
$$
G^L=\left[\begin{array}{cc}0&D^2\\
0&0\\
3D^4&D^2
\end{array}\right]=
\left[\begin{array}{cc}0&1\\
0&0\\
3&1
\end{array}\right]\left[\begin{array}{cc}
D^4&0\\
0&D^2
\end{array}\right].
$$
Therefore, the above corollary  should be regarded as a generalization of the classical Forney's theorem stating that a full column rank polynomial
matrix has the PD property if and only if its leading coefficient matrix has full column rank.

\section{Minimal reduced polynomial resolutions, and Forney tables}

A priori is not clear that every convolutional code possesses  reduced polynomial resolutions.
The issue of minimality also is not obvious.
(In dimension 1, reduced polynomial matrices automatically are minimal.)
 In this section, we shall see that for every convolutional code there exists a minimal reduced polynomial resolution and that such a resolution is uniquely determined up to equivalence.

Let $(M,(M_{\leq d}))$ be a filtered module. If $d\geq 0$, then
$$M_{\leq d-1}+D_1M_{\leq d-1}+\cdots +D_nM_{\leq d-1}$$ is the part of $M_{\leq d}$ that comes from $M_{\leq d-1}$ and
  should be regarded as the trivial part of $M_{\leq d}$. It is natural therefore to consider the quotient
$$
 \frac{M_{\leq d}}{M_{\leq d-1}+D_1M_{\leq d-1}+\cdots +D_nM_{\leq d-1}},
$$
which is a linear space over $\mathbb{F}$. Denote it by $\Gamma_d(M,(M_{\leq d}))$.

{\em Example}. There holds
$$ \Gamma_d(S[-k])=
\left\{\begin{array}{cc} \mathbb{F} & \ \tx{when} \ d=k;\\
\{0\} & \ \tx{when} \ d\neq k.
\end{array}\right.
$$

Let $u:(M,(M_{\leq d}))\ra (N,(N_{\leq d}))$ be a homomorphism of filtered modules. We say that $u$ is minimal if
it satisfies the following two conditions:

 1) $u: M_{\leq d} \ra N_{\leq d}$\ is surjective \ $\forall d\geq 0$;

 2) $\Gamma_d(u):\Gamma_d(M,(M_{\leq d}))\ra \Gamma_d(N,(N_{\leq d}))$ \ is bijective \ $\forall d\geq 0$.

\begin{lem} Let $p\geq 1$, and let  $C$ be a (nontrivial) convolutional code in $S^p$.
For any twisting function $a: [1, p]\ra \mathbb{Z}_+$,
 there exists a  polynomial matrix   $G$ such that
$$G:S^q[-b]\ra C[-a],$$
where $q$ is the column number and $b$ the column $a$-degree of $G$,  is minimal.
\end{lem}

{\em Proof}. See Lemma 7 in \cite{L2}.
 $\quad\Box$

Any $G$ satisfying the conditions of the lemma is called a minimal $a$-representation of $C$. In the case when
$a=0$, we simply say "minimal representation".

An exact sequence of filtered modules
$$
(F_l,(F_{l,\leq d}))\stackrel{\delta_l}{\ra} \cdots \stackrel{\delta_2}{\ra}
  (F_1,(F_{1,\leq d}))
\stackrel{\delta_1}{\ra} (F_0,(F_{0,\leq d}))
$$
is said to be minimal if $\delta_1$ is minimal and, for each $i\geq 2$, the homomorphism
$$
\delta_i: (F_i,(F_{i,\leq d}))\ra (Ker\delta_{i-1},(Ker\delta_{i-1}\cap F_{i-1,\leq d}))
$$
also
is minimal.

{\bf Definition.}\
Let $G=(G_1,\ldots, G_l)$ be a reduced polynomial resolution, and let $q\times (p_l,\ldots ,p_1)$ be its size
and $(a_l,\ldots ,a_1)$ the column degree. Say that $G$ is minimal if the sequence (1) of filtered modules
is minimal.

One constructs a  minimal reduced polynomial resolution step by step, using the previous lemma.
Assume  we have a convolutional code $C\subseteq S^q$.
 Choose a minimal representation $G_1$ of $C$. Let $C_1$  be the kernel of $G_1:S^{p_1}\ra C$, where $p_1$ is the column number of $G_1$.
 Letting $a_1$ denote the degree function of $G_1$,  choose next a minimal $a_1$-representation $G_2$ of $C_1$. If we continue this way the process will terminate.

\begin{thm} Every convolutional code  has a minimal reduced polynomial resolution.
\end{thm}

{\em Proof}. This is an immediate consequence of the "graded" Hilbert syzygy theorem. (See Theorem 4.15 in Lang \cite{L} and Theorem 1 in \cite{L2}.) $\quad\Box$

Let
 $G=(G_l,\ldots , G_1)$ and $G^\prime=(G^\prime_l,\ldots , G^\prime_1)$  be two reduced polynomial resolutions. Let $q\times (p_l,\ldots ,p_1)$ and $q\times (p^\prime_l,\ldots ,p^\prime_1)$ be their sizes, and let
  $(a_l, \ldots ,a_1)$ and $(a^\prime_l, \ldots ,a^\prime_1)$ be their
 column degree tables. We say that $G$ and $G^\prime$  are
  equivalent if there exist  isomorphisms
$$ U_i: S^{p_i}[-a_i] \ra S^{p^\prime_i}[-a^\prime_i], \ \ \ \ i=1,\ldots , l$$
such that
$$ G^\prime_1U_1=G_1, \ G^\prime_2U_2=U_1G_2, \ldots ,\ G^\prime_lU_l=U_{l-1}G_l.$$

It is immediate from Lemma 1 that
 equivalent reduced polynomial resolutions have the same size and the same column degree table.

\begin{thm} Any two minimal reduced polynomial resolutions of a convolutional code are  equivalent.
\end{thm}

{\em Proof}. This is an immediate consequence of the theorem about uniqueness of minimal graded free resolutions.
(See Theorem 6.3.13 in Cox,  Little and O'Shea \cite{CLS},  Theorem 1.6  in  Eisenbud \cite{E2}, and  Theorem 1 in \cite{L2}.) $\quad\Box$

The above two theorems permit us to give the following definition.

{\bf Definition.}\
 Let $C\subseteq S^q$ be a convolutional code, and let
 $G=(G_l,\ldots , G_1)$ be any its minimal reduced polynomial resolution. We define the rate
 of $C$ to be $(p_l,\ldots ,p_1)/q$, where  $(p_l,\ldots ,p_1)$ is the size of $G$. Next, we define
  the Forney table of $C$ to be the column degree table of $G$.
The maximum value of the degree function of $G_1$ is called the {\em memory} of the code.

\begin{prop} Let $C$ be a convolutional code, and let $m$ be its memory. Then $C$ can be recovered from the knowledge of \ $C_{\leq m}$.
\end{prop}

{\em Proof}. Let $G_1$ be a minimal representation of $C$, and let $p_1$ its column number, $a_1$ the degree function and $m$ the memory. For $d> m$, we have
$$
\Gamma_d(S^{p_1}[-a_1])=0.
$$
Because $G_1: S^{p_1}[-a_1]\ra C$ is minimal,
we get that $\Gamma_d(C)=0$ for all $d> m$. In other words, for all such $d$, we have
$$
C_{\leq d}=C_{\leq d-1}+D_1C_{\leq d-1}+\cdots +D_nC_{\leq d-1}.
$$
It follows that  knowing $C_{\leq m}$, we can find all $C_{\leq d}$ with $d> m$. This completes the proof since, for any $N$,
$$C=\bigcup_{d\geq N} C_{\leq d}.$$
$\quad\Box$

Closing the section, we want to present
 a simple test for  establishing whether a given reduced polynomial resolution is minimal or not.

\begin{prop} Let   $G=(G_l,\ldots , G_1)$ be a  reduced polynomial resolution. If $l=1$, then $G$ is minimal.
If $l\geq 2$, the $G$ is minimal if and only if none of the nonzero entries of the matrices
$G_2^L, \ldots , G_l^L$ are scalars from $\mathbb{F}$.
\end{prop}

{\em Proof}. The case $l=1$ is obvious. The resolution $G$ is minimal if and only if the sequence of graded free modules (3) is minimal.
(The reader is referred to Eisenbud \cite{E2} for the notion of minimal graded free resolutions.) By
 Corollary 1.5 in \cite{E2},
 (3) is minimal if and only if the scalar matrices
$$
G_k^H(0,0,\ldots,0), \ \ \ \ k=2, \ldots , l
$$
are zero. This completes the proof since
$$
G_k^L(0,\ldots,0)=G_k^H(0,0,\ldots,0).
$$
 $\quad\Box$

\section{Observability}

A desirable property from the point of view of coding is observability.

A convolutional code
$C\subseteq S^q$
is called
observable
if the quotient module $S^q/C$ is torsion free.  As is known, a finitely generated module is torsion free if and
only if it can be embedded into a free module of finite rank. It follows that
 $C$ is observable if and only if it can be described through
$$
C=\{f\in S^q|\ Hf=0\},
$$
where $H\in S^{\bullet\times q}$ is a polynomial matrix, called a parity check matrix or a syndrome former of $C$.

 For an irreducible polynomial $\lambda$ in $S=\mathbb{F}[D]$, let $\mathbb{F}(\lambda)$ denote the integral domain
   $\mathbb{F}[s]/\lambda\mathbb{F}[s]$. If $G$ is a polynomial matrix, define $G/\lambda$ to be its reduction modulo
   the principal ideal $\lambda\mathbb{F}[s]$.

   We   have the following proposition.

\begin{prop} Let $C\subseteq S^q$ be a convolutional code, and let $(G_l,\ldots ,G_1)$ be its
polynomial resolution of size $(p_l,\ldots, p_1)$, say. Then $C$ is observable if and only if the sequence
$$
0\ra \mathbb{F}(\lambda)^{p_l}\stackrel{G_l/\lambda}{\ra}
 \cdots \ra
\mathbb{F}(\lambda)^{p_1}\stackrel{G_1/ \lambda}{\ra} \mathbb{F}(\lambda)^q
$$
is exact for every irreducible polynomial $\lambda$.
\end{prop}

{\em Proof}. See Theorem 1 in  \cite{L1}.
 $\quad\Box$

{\em Remark}. In dimension 1, the statement is well-known in linear systems theory, where it is named as the
Popov-Belevich-Hautus test of controllability.


\begin{thebibliography}{99}


\bibitem{C} C. Charoenlarpnopparut, "Applications of Gr${\rm{\ddot{o}}}$bner bases to the structural
description and realization of multidimensional convolutional code", {\em Science Asia}, vol. 35, pp. 95-105, 2009.

\bibitem{CLS} D.A. Cox, J. Little and D. O'Shea,  {\em Using Algebraic Geometry}. New York, Springer-Verlag,  2005.

\bibitem{E2} D. Eisenbud,  {\em The Geometry of Syzygies: A Second Course in Algebraic Geometry and Commutative Algebra}. New York, Springer-Verlag, 2005.

\bibitem{FV} E. Fornasini and M.E. Valcher, "Algebraic aspects of two-dimensional convolutional codes", {\em IEEE
Trans. Inform. Theory}, vol. 40, pp.  1068-1082, 1994.

\bibitem{F} G.D. Forney Jr., "Convolutional codes I: Algebraic structure", {\em IEEE Trans. Inform. Theory}, vol. IT-16, pp. 720-738, 1970.

\bibitem{GRW} H. Gluesing-Luerssen, J. Rosenthal and P.  Weiner, "Duality between multidimensional convolutional codes and systems", in
{\em Advances in mathematical systems theory}, F. Colonius, U. Helmke, D. Prätzel-Wolters and F. Wirth, Eds. Boston, MA, 2001, pp. 135-150.

\bibitem{JC} P. Jangisarakul and C. Charoenlarpnopparut, Algebraic decoder of a multidimensional convolutional code: constructive algorithms for determining syndrome decoder and decoder matrix based on  Gr${\rm{\ddot{o}}}$bner basis, {\em Multidim. Syst.
Sign. Process.}, vol. 22. pp. 67-81, 2011.

\bibitem{K} B. Kitchens, "Multidimensional convolutional codes", {\em SIAM J. Discrete Math}., vol. 15, pp.  367-381, 2002.

\bibitem{L} S. Lang, {\em Algebra}. New York, Springer-Verlag,  2002.

\bibitem{L1} V. Lomadze,  "PBH test for multivariate LTID systems", {\em Automatica}, vol. 49, pp. 2933-2937, 2013.

\bibitem{L2} V. Lomadze, ""Reduced polynomial matrices" in several variables", {\em SIAM J. Control Optim.},  vol. 51, pp. 3258-3273, 2013.

\bibitem{MS} R.J. McEliece and R.P. Stanley,   "The general theory of convolutional codes",
{\em The Telecommunications and Data Acquisition Progress Report 42-113}, pp. 89-98, January-March 1993.

\bibitem{NPP1} D. Napp Avelli, C. Perea and R. Pinto, "Input-state-output representations and constructions of finite-support 2D convolutional codes", {\em Advances in Mathematics of Communications}, vol. 4, pp. 533-545, 2010.

\bibitem{NPP2} D. Napp Avelli, C. Perea and R. Pinto, "Column distances for 2D-convolutional codes",
{\em Proceedings of the 19th International Symposium on MTNS},  Budapest, Hungary, 2010.

\bibitem{VF} M.E. Valcher and E. Fornasini. "On 2D finite support convolutional codes: an algebraic approach",
{\em Multidim. Syst. Sign. Process.}, vol. 5, pp. 231-243, 1994.

\bibitem{W} P. Weiner. "Multidimensional Convolutional Codes", Ph.D. dissertation,
University of Notre Dame, USA, 1998.

\bibitem{Z} E. Zerz, "On multidimensional convolutional codes and controllability properties of multidimensional systems over finite rings", {\em Asian Journal of Control}, vol. 12, pp.  117-236, 2010.

 \end{thebibliography}
\end{document}